\documentclass[12pt,preprint]{aastex}

\slugcomment{Submitted to The Astrophysical Journal}

\shorttitle{Spherical MHD Accretion}
\shortauthors{I. V. Igumenshchev}

\begin{document}

\title{Three-Dimensional Simulations of Spherical Accretion Flows\\
with Small-Scale Magnetic Fields}

\author{Igor V. Igumenshchev}
\affil{Laboratory for Laser Energetics, University of Rochester\\
250 East River Road, Rochester, NY 14623}
\email{iigu@lle.rochester.edu}

\begin{abstract} 

Spherical (nonrotating) accretion flows 
with small-scale magnetic fields have been investigated
using three-dimensional, time-dependent MHD simulations.
These simulations have been designed to model high-resolution (quasi) steady 
accretion flows in a wedge computational domain 
that represents a small fraction of the full spherical domain. 
Subsonic and supersonic (super-fast-magnetosonic)
accretion flows have been considered.
Two accretion regimes have been studied: conservative, or 
radiatively inefficient, and nonconservative, in which the heat
released in magnetic reconnections is completely lost.
The flows in both regimes are turbulent.
They show the flattened radial density profiles and
reduction of the accretion velocities and mass accretion rates
in comparison with hydrodynamic Bondi flows.
In the conservative regime, the turbulence is 
more intensive and supported mostly by thermal convection.
In the nonconservative regime, the turbulence is
less intensive and supported by magnetic buoyancy and
various magnetic interactions.
We have concluded that steady, supersonic spherical accretion 
cannot be developed in the presence of small-scale magnetic fields.

\end{abstract}


\section{Introduction}

Mass accretion onto a central gravitational body plays an
important role in the formation and evolution of a large variety of
astrophysical objects such as planets, stars, 
galactic nuclei, galaxies, and clusters of galaxies 
(see Frank, King \& Raine 1992). 
Depending on the amount of the angular momentum $\ell_0$, which
the accreting mass carries at the outer boundary radius $R_{\rm out}$,
accretion flows can take either a disk-like or spherical-like form.
Flows with relatively high angular momentum,
$\ell_0\simeq\sqrt{GMR_{\rm out}}$, where $G$ is the gravitational
constant and $M$ is the central mass,
form centrifugally supported accretion disks (e.g., Shakura 1972).
Flows with low angular momentum, $\ell_0\ll\sqrt{GMR_{\rm out}}$,
can form spherical or quasi-spherical accretion flows at radii 
$R\ga \ell_0^2/GM$ in which the centrifugal force is weak and
not sufficient to balance gravity (Illarionov \& Sunyaev 1975).
The theoretical study of spherical accretion flows is based on 
an analytical solution discovered several decades ago
by Bondi (1952). This solution describes
the idealized case of isentropic nonmagnetic accretion flows. 
Since that time, the theory of spherical accretion has been
significantly developed so that it now includes an understanding 
the role of different
physical mechanisms (such as radiative cooling and heating, magnetic field
dissipation, thermal and radiation transports, etc.) 
and more realistic inner and outer boundary conditions 
(for reviews, see Frank {\it et al.} 1992; Kato, Fukue, \& Mineshige 1998).
More recent studies of spherical accretion
considered convection (Markovi\'c 1995),
accretion onto a magnetic dipole (Toropin {\it et al.} 1999),
magnetic diffusivity due to turbulence (Shadmehri 2004), and vorticity
(Krumholz, McKee, \& Klein 2005).

This paper continues the numerical
study of radiatively inefficient spherical (nonrotating)
accretion flows with magnetic fields
conducted by Igumenshchev \& Narayan (2002, hereafter IN; also
see Pen, Matzner, \& Wong 2003 for a similar study). 
IN have demonstrated, with help of three-dimensional MHD
simulations, that the effects of a magnetic field can significantly
modify the structure of Bondi-type flows. They argued that
even initially weak magnetic fields can produce dramatic changes.
The main reason for these changes is the local, 
nonuniform release of
the thermal energy during the dissipation of tangled magnetic fields
in reconnections. This release suppresses
the inward motion of mass and results in the development of turbulence
that is mainly supported by thermal convection.
The dissipation of a magnetic field in reconnections 
is compensated by the efficient
amplification of the radial field component in spherical convergent
flows (Shvartsman 1971).
IN have developed a simple analytical theory of 
spherical convection-dominated accretion flows 
(spherical CDAFs\footnote{Note that IN used instead the name
``convection-dominated Bondi flows" which we do not adopt here.}),
which is similar, with regards to the involved basic physics,
to the theory of rotating CDAFs 
(Narayan, Igumenshchev, \& Abramowicz 2000; Quataert \& Gruzinov 2000).
This theory predicts the flattened radial density profile
$\rho\propto R^{-1/2}$ in contrast to the steeper density profile
in asymptotic Bondi flows, $\rho\propto R^{-3/2}$ 
(see the appendix, eq.~[A6]).
Because of this flattened profile, the mass accretion rate
in spherical CDAFs is expected to be significantly lower than
the Bondi mass accretion rate in flows with the 
equivalent outer boundary conditions;
\begin{equation}
\dot{M}\sim\dot{M}_{\rm Bondi}\left({R_{\rm in}\over R_{\rm out}}\right),
\end{equation}
where $R_{\rm in}$ is the inner radius of the flows.
These properties make the spherical CDAF solution
to be a prominent candidate for explaining
the phenomenon of dim galactic nuclei contained supermassive black
holes, including our Galactic center Sgr A$^*$
(e.g., Melia 1992; Baganoff {\it et al.} 2001, 2003; Quataert 2002; 
Ghez {\it et al.} 2003; Ho, Terashima, \& Okajima 2003; 
Soria {\it et al.} 2006), 
in which accretion disks are typically not observed.
Also, spherical CDAFs can be employed to explain the problem of
missing isolated neutron stars in our Galaxy 
(Treves \& Colpi 1991; Blaes \& Madau 1993; Turolla {\it et al.} 1994;
Belloni, Zampieri, \& Campana 1997; 
Toropina {\it et al.} 2003; Perna {\it et al.} 2003) 
and the observed properties of isolated stellar mass black holes 
(e.g., Fujita {\it et al.} 1998; Agol \& Kamionkowski 2002).

The generality of the numerical results of IN and Pen {\it et al.} (2003) 
were limited in the important aspect that no steady or quasi-steady
accretion flows were obtained.
These authors studied transient states of the flows, which originated
because of specific initial conditions and involved moving outward shocks.
These results also suffered from insufficient numerical resolution,
especially in the innermost region of the flows.
In addition, the assumption of an initial bipolar magnetic field
in IN resulted in the domination of a large-scale poloidal 
field and doughnut-like density distribution 
near the black hole at the later evolution time 
(note that similar field topology was proposed by
Bisnovatyi-Kogan \& Ruzmaikin 1974). Unless such large-scale
poloidal field can be naturally present in some objects,
other possible field configurations, such as a small-scale field with zero net
magnetic flux, can likely be natural too in spherical accretion flows.
The main goal of the present study is the investigation of the role
of small-scale magnetic fields in such flows.

In the numerical aspect, we overcome some of the
limitations of the IN's approach
employing a new simulation design. In this design, we assume
a permanent injection of mass and magnetic field into the computational domain
that allows us to obtain steady or quasi-steady accretion flows
after performing long-time evolution simulations.
The numerical resolution
has been improved by adopting spherical coordinates and
conducting simulations in the wedge computational domain, which
represents a small fraction of the full spherical domain (see Fig.~1).
The injected field is assumed to have
a small-scale component in the form of radially extended 
magnetic loops with a zero net magnetic radial flux (see Fig.~2).
Modifications of the simulation technique allow us to
investigate two limiting energetic regimes: conservative (or
radiatively inefficient) and non-conservative (in which the heat
from magnetic reconnections is completely lost).

The paper is organized as follows.
In Section 2 we describe the simulation technique, 
initial and boundary conditions,
and algorithm of the mass and magnetic field injection. 
Section 3 presents numerical results, and
in Section 4 we discuss these results and make final conclusions.
We reproduce some analytic solutions
of hydrodynamic accretion flows, including Bondi solution,
in the appendix.

\section{Simulation Technique}

We use the numerical method, which is similar to that used by IN
and Igumenshchev, Narayan, \& Abramowicz (2003).
The method solves the system of
ideal MHD equations (e.g., Landau \& Lifshitz 1987), which
describe the dynamics of nonself-gravitating plasmas 
in the central gravitational field.
Originally, the method employed a nonconservative numerical scheme
that solves the internal energy equation, which includes the
reconnection heat term $Q$ (for more details see IN);
\begin{equation}
\rho{d\epsilon\over dt} = -P_{\rm g}{\bf\nabla\cdot v} + Q,
\end{equation}
where $\rho$ is the density, $\epsilon$ is the specific internal energy,
$P_{\rm g}$ is the gas pressure, and ${\bf v}$ is the velocity.
Test simulations have shown that MHD solutions obtained using 
equation (2) conserve the total energy quite poorly because
of artificial loosing or gaining energy in numerical reconnections. 
In some our test cases, the relative error of the total energy 
conservation was up to 10\%.
To solve this problem, the method has been modified by adding a
conservative option. 
Using this option, the method solves the total energy equation
\begin{equation}
{\partial\over\partial t}\left(\rho{v^2\over 2}+\rho\epsilon+
{B^2\over 8\pi}\right)=-\nabla\cdot{\bf q}
\end{equation}
instead of equation (2).
Here ${\bf B}$ is the magnetic induction and
${\bf q}$ is the total energy flux per unit square.
Note that in finite-difference MHD schemes the magnetic field is
reconnected on scales in which the minimum is limited by the gridsize. 
Typically, this gridsize
is much larger than the physical reconnection scales in the studied problems.
Therefore, the finite-difference schemes, including our
scheme, can not accurately represent all the details in the process of
reconnection. For our purposes, however, the provided level of accuracy
is sufficient.
This situation is somewhat analogous to the representation
of shocks in finite-difference hydrodynamic schemes in which the
numerical shock thickness is also limited by the gridsize.

In all our simulations, we have used the ideal gas equation of state
\begin{equation}
P_{\rm g}=(\gamma-1)\rho\epsilon
\end{equation}
and assumed the adiabatic index $\gamma=5/3$.

This method employs the three-dimensional
spherical coordinates ($R,\theta,\phi$).
The computational domain is limited by a narrow four-facet wedge
located at the equatorial plane as shown in Figure~1. 
The domain is extended from $R_{\rm in}$ to $R_{\rm out}$
in the radial direction and over $\theta_0$ and $\phi_0$ degrees in
the polar and azimuthal directions. We have assumed
$\theta_0=\phi_0=\pi/16$ and the number of angular gridpoints 
$N_\theta\times N_\phi=30\times 30$. These points are uniformly spaced in
both $\theta$ and $\phi$ directions.
Gridpoints in the radial direction are spaced logarithmically so that the
three-dimensional numerical cells at any radius take an approximately 
cubic shape. This provides the direction-independent 
local spatial resolution in the simulations.
The number of the radial gridpoints $N_R=303$, which corresponds to
$R_{\rm out}/R_{\rm in}=10$.

\subsection{Boundary Conditions}

We have used three different sets of boundary conditions 
assumed at the azimuthal, polar, and radial boundaries, respectively,
of the wedge computational domain (see Fig.~1).
The periodic boundary conditions for both fluid and magnetic field
are applied at the azimuthal boundaries.
At the polar boundaries, we use the conditions that no streamlines 
and magnetic lines can go inside/outside through these boundaries.
This is achieved by applying the reflection boundary conditions
for fluid and assuming the
continuous tangential and reflection normal magnetic components across
the polar boundaries.

At the inner $R_{\rm in}$ and outer $R_{\rm out}$ radial boundaries,
we apply the absorbing boundary conditions for a fluid.
This means that the mass can flow
freely out of the computational domain, but no mass is allowed to
return from outside.
Conditions for the magnetic field are assumed considering
``ghost" boundary zones located on the outside of the computational domain.
We assume that these zones can contain only a radial magnetic field,
whose strength is determined from the divergence-free constraint
$\nabla\cdot{\bf B}=0$.
Numerical tests have shown that a mass with a frozen-in tangled magnetic 
field can freely flow out through these radial boundaries
without the effects of field or mass accumulation.

Our absorbing boundary conditions at $R_{\rm in}$ qualitatively
correctly mimic the conditions near the black hole horizon where
matter is free falling in the strong gravity field and 
the radial component of the magnetic field dominates the tangential component
(e.g., Thorne, Price, \& MacDonald 1986).
In the case of accreting stars with rigid surfaces such as
white dwarfs and neutron stars, which can also have magnetospheres,
the inner boundary condition will depend on the radiative efficiency
of the flows. Radiatively inefficient plasma will probably form
a slowly accreting ($\dot{M}\ll\dot{M}_{\rm Bondi}$)
subsonic accretion flows, similar to that
``a tenuous continuation of the star" discussed by Bondi (1952).
If plasma efficiently radiates its thermal energy near the stars'
surfaces (e.g., Shapiro \& Salpeter 1975), 
the considered absorbed inner boundary conditions
can be adequate for accretion flows far away from these surfaces.

\subsection{Injection of Mass}

To obtain steady, or quasi-steady, accretion flows,
we permanently inject mass into the computational domain.
The main problem here is to minimize the consequences of 
the interaction of outflows, which can originate during simulations,
and the injected mass.
We have employed an injection algorithm, which has
some resemblance to the injection
algorithm used by Igumenshchev \& Abramowicz (1999) in simulations
of rotating accretion flows.

We assume that the mass is steadily injected
inside a thin spherical layer with the radius $R_{\rm inj}$,
which is located closely to the outer absorbing boundary at $R_{\rm out}$.
This mass is distributed uniformly over $\theta$ and
$\phi$ directions and has zero velocity.
Under the action of gravity, the larger fraction of the
injected mass forms an accretion flow.
The smaller fraction of this mass could
escape through $R_{\rm out}$
because of a thermal spread and interactions with outflows.

The injected mass is characterized by the specific
internal energy $\epsilon_0$, which determines two regimes of
hydrodynamic accretion -- subsonic and supersonic.
The critical value
$(\epsilon_0)_{\rm crit}\approx\epsilon_{\rm vir}$ where
\begin{equation}
\epsilon_{\rm vir}\equiv GM/R_{\rm inj}
\end{equation}
separates these regimes:
$\epsilon > (\epsilon_0)_{\rm crit}$ corresponds to subsonic flows and
$\epsilon < (\epsilon_0)_{\rm crit}$ corresponds to supersonic flows
(see the appendix).

\subsection{Injection of a Magnetic Field}

The injected mass can also carry a magnetic field that is associated with it. 
The field is injected assuming
that only the poloidal component $A_\theta$ of the vector potential
${\bf A}$ is nonzero in the injected mass; 
the other two components, $A_R$ and $A_\phi$, are set to zero. 
In this way we can produce
$B_R$ and $B_\phi$ components, which obey
\begin{equation}
B_R=-{1\over R \sin\theta}{\partial A_\theta\over\partial\phi}, \qquad
B_\phi={1\over R}{\partial \over\partial R}\left(RA_\theta\right).
\end{equation}

Three different magnetic field configurations have been used
in simulations: purely (unipolar) radial, purely toroidal, and loop
configurations.
The former two configurations have been chosen 
to test our numerical method.
The purely radial field
is generated at the beginning of simulations assuming that
\begin{equation}
A_\theta=B_{\rm inj}R_{\rm inj}{R_{\rm inj}\over R} \phi,
\end{equation}
where $B_{\rm inj}$ is the magnetic induction at $R_{\rm inj}$.
This field remains unchanged during simulations
by virtue of the conservation of the radial magnetic flux confined in the
wedge computational domain.

Two other field configurations, purely toroidal and loop fields, 
are formed by permanently injecting the corresponding field at $R_{\rm inj}$. 
This field injection is tightened to the mass injection
and performed by correcting $A_\theta$ inside the injection layer
at each time step as follows:
\begin{equation}
A_\theta'=A_\theta+\xi \Delta A,
\end{equation}
where $\Delta A$ is an increment 
and $\xi$ is a correction factor $0\le\xi\le 1$.
The increment $\Delta A$ depends on the assumed field configuration.
In the case of purely toroidal field,
\begin{equation}
\Delta A=B_0 R_{\rm inj},
\end{equation}
where $B_0=\sqrt{8\pi\gamma (\gamma -1)\epsilon_0\Delta\rho/\beta_0}$ 
and $\Delta\rho$ is the density of the injected mass.
In the case of loop field,
\begin{equation}
\Delta A=B_0 R_{\rm inj}{R_{\rm inj}\over R} {\sin(m\phi)\over m},
\end{equation}
where the parameter $m$ determines the number of
azimuthal sectors in which the radial magnetic field 
periodically changes direction. A combination of two such 
sectors forms a magnetic loop.
In the present simulations, we have assumed two magnetic loops
in the computational domain as illustrated in Figure~2.

The correction factor $\xi$ in equation (8) is used to maintain 
the strength of the injected field
at a given level determined in terms of the plasma 
$\beta\equiv P_{\rm g}/P_{\rm m}$,
where $P_{\rm m}=B^2/8\pi$ is the magnetic pressure. We assume that
\begin{equation}
\xi={\rm min}\left[1,{\rm max}\left(0,
  {\langle\beta\rangle-\beta_0\over\beta_0}\right)\right],
\end{equation}
where $\beta_0$ is a parameter, which determines the field strength,
and $\langle\beta\rangle$ is the $\beta$
averaged over the volume of the injection layer.

\section{Numerical Results}

We initiate simulations from a static, nonmagnetic,
very low-density medium that fills the entire computational domain.
Accretion flows are created by steadily injecting the mass
into this medium at $R_{\rm inj}$.
These flows generally go
through an initial transient phase before relaxing to a steady 
or quasi-steady state.
The transient phase typically takes a few tens ($\simeq 10-30$)
of the free-fall time, 
$t_{\rm ff} \approx 0.67 R_{\rm out}^{3/2}/\sqrt{GM}$, and
includes the formation and dissipation of shocks and nonlinear waves. 
The hydrodynamic flows and
MHD flows with the purely radial and toroidal fields have 
finally been relaxed to steady states.
The MHD flows with the loop field have remained 
time variable, or quasi-steady, even after the completion of 
the transient phase.

IN discussed in detail the role of magnetic reconnections
in spherical accretion flows with $\beta\sim 1$.
This role depends on the amount of reconnection heat 
contributed to the gas internal energy.
We have considered two extreme energetic regimes,
conservative and nonconservative.
In the conservative regime, the dissipated magnetic energy is
fully transformed into the heat.
Models in this regime have been calculated by
employing the total energy equation (3).
In the nonconservative regime, the dissipated magnetic
energy is totally lost and does not produce any heat.
In this regime, we employ the internal energy equation (2) 
in which the dissipation term $Q$ has been set to zero.
Note that the term descibed the artificial heat from shocks
has been retained in equation (2).

The study of these conservative/nonconservative accretion regimes 
can be astrophysically motivated. The conservative regime can
corresponds to the very high accretion rate flows with
$\dot{M}\gg\dot{M}_{\rm Edd}$, which do not
radiate much energy due to the large optical depth of the flows
(e.g., Katz 1977).
Here $\dot{M}_{\rm Edd}=4\pi GM/\sigma_{T}m_p c$ 
is the Eddington accretion rate, $\sigma_{T}$ is the Thomson-scattering
cross-section, and $m_p$ is the proton mass.
Another option for the conservative regime is the 
very low accretion rate flows, $\dot{M}\ll\dot{M}_{\rm Edd}$, 
if one assumes that the energy released
in magnetic reconnections primary does to the ions, which cannot
loose this energy efficiently via the usually considered
electron-ion Coulomb collisions
because of the tenuous plasma (e.g., Esin {\it et al.}).
There is another possibility, however,
that a significant fraction of the
reconnection energy will go to electrons, which can
efficiently radiate this energy by variety of mechanisms
(Bisnovatyi-Kogan \& Lovelace 1997, 2000; Quataert \& Gruzinov 1999).
The latter possibility provides the additional cooling mechanism for ions, 
which in some circumstance could be
more efficient than the Coulomb collisions,
making the low, or even very low, accretion rate flows to be nonconservative,
or radiatively efficient.

We shall describe models that are either subsonic or supersonic and
differed by the strength of
the injected fields, amount of the heat release in reconnections,
and field configuration.
The model parameters are listed in Table~1.
For convenience, we have divided all these models into three groups:
Bondi-type flows (Models~A1-A4), subsonic MHD flows (Models~B1 and B2),
and supersonic MHD flows (Models~C1 and C2). The structure of 
the Bondi-type flows, i.e. flows without magnetic fields or with the fields 
of special topology, 
is similar to the structure of Bondi (1952) solution and
well approximated by nonmagnetic analytic solutions discussed in the appendix.
The structure of the latter two groups of models is significantly
modified because of magnetic fields.
%not affected by magnetic fields.
We shall describe these groups separately.

\subsection{Bondi-type flows}

Bondi-type flows (Models~A1-A4, see Table~1)
have been calculated to verify our numerical method and 
obtain reference models.
These flows are steady and laminar.
Models~A1-A3 represent subsonic flows and Model~A4 represents 
a supersonic flow.
Models~A1 and A4 have no magnetic field, and thus describe
hydrodynamic flows.
Model~A2 has the purely toroidal but
sufficiently subequipartition ($\beta\gg 1$)
magnetic field and  Model~A3 has the uniformly ordered strong ($\beta\la 1$) 
radial magnetic field. In the latter two cases, the magnetic field
does not affect the structure and dynamics of the flows and
these cases are almost equivalent to Model~A1.

We shall describe the evolution and structure of subsonic Bondi-type flows 
using Model~A1 as a representative example.
This model has evolved from the initial state (see \S2)
to a steady state through the transient phase.
This phase includes the formation of temporal shocks and waves, which
many times propagate inward and outward in the radial direction, reflecting
from the inner and outer boundaries, before
complete disappearance after $t\simeq 20-30 t_{\rm ff}$.
The final flow in Model~A1 is steady, effectively
one-dimensional, and described by the analytic solution 
(eqs.~[A4] and [A8]). This solution suggests that Model~A1 
represents a part of the flow located deeply inside the Bondi radius,
$R_{\rm out}=0.0371 R_B$, and,
therefore, this model can be closely approximated by an
asymptotic Bondi solution (A6), which appears in the limit $R\ll R_B$.
Figure~3 shows selected properties of Model~A1 (short-dashed lines)
as functions of the radius. These properties
include the distribution of density $\rho$ (left top panel); 
gas pressure $P_{\rm g}$ (left middle panel); relative temperature
$T/T_{\rm vir}=\epsilon/(GM/R)$ (left bottom panel); 
Mach number ${\cal M}=v/c_s$ (right top panel),
where $c_{\rm s}=\sqrt{\gamma P/\rho}$ is the sound speed; relative
radial velocity $v/v_{\rm ff}$ (right middle panel); 
and relative accretion rate (right bottom panel). 
In the latter case, the accretion rate
is normalized to the mass injection rate (see \S2).
Because the flow is steady, the accretion rate is independent of the radius.
Note that about 90\% of the injected mass forms inflow.
The other 10\% escapes through $R_{\rm out}$ 
because of a thermal spread.

For comparison, Figure~3 also shows a self-similar solution (A9) 
in which ${\cal M}=1$ (long-dashed lines).
This solution is ``boundary free" and
has asymptotic Bondi profiles
$\rho\propto R^{-3/2}$ and $P_{\rm g}\propto R^{-5/2}$ throughout.
Note that Model~A1 demonstrates slightly flatter density and pressure
profiles, which
can be explained by the deviation of $R/R_B$ from the zerolimit.

Model~A2 has the weak and dynamically unimportant toroidal magnetic field
$\beta\gg 1$. In all other aspects, Models~A2 and A1 are similar.
Using Model~A2, we have tested the ability of our numerical method to
accurately model the passive transport of magnetic field.
The magnetic flux conservation requires that the toroidal field near the 
equatorial part of a spherical flow is changed as 
\begin{equation}
B_\phi\propto (R v)^{-1}.
\end{equation}
This leads to $B_\phi\propto R^{-1/2}$ and $P_{\rm m}\propto R^{-1}$
for the self-similar velocity profile $v\propto R^{-1/2}$. 
Because Model~A2 has the steeper velocity profile, however,
the actual profile of $P_{\rm m}$ obtained in this model is accordingly 
flatter and its slope is fully consistent with estimate (12).

Model~A3 has a uniform radial magnetic field. This field is changed
with the radius as $B_R\propto R^{-2}$ and therefore
$P_{\rm m}\propto R^{-4}$.
The strength of the field has been chosen such that it has $\beta = 30$
at $R_{\rm out}$ and $\beta \simeq 0.1$ at $R_{\rm in}$.
The simulations have shown no effects from the flow transition
from the subequipartition to superequipartition field regions.
We shall see later, however, that such a transition causes significant
changes in the flows with the loop field.

Note the artificial
nature of the field topology assumed in Model~A3 that
represents a small sector of the spherically symmetric monopole field.
Such a field can not have magnetic reconnections, which
actually play an important role in realistic MHD flows.
Accretion flows in a strong unipolar magnetic field 
similar to that considered in Model~A3, however,
can occur at magnetic poles of neutron stars and white dwarfs.

Properties of the supersonic hydrodynamic Model~A4 are shown in
Figure~4 with short-dashed lines. The density and pressure profiles
in this model demonstrate nonmonotonic behavior 
at the outer region and approaching the asymptotic Bondi powerlaws
$\rho\propto R^{-3/2}$ and $P_{\rm g}\propto R^{-5/2}$ at the inner region
(see left top and left middle panels in Fig.~4). 
The Mach number and accretion velocity
(see right top and right middle panels in Fig.~4)
show a significant increase inward from the outer boundary.
All these properties of Model~A4 are in good agreement with
analytic solution (A11), which fits Model~A4, 
assuming $R_{\rm out}/\tilde{R}_B=0.071$ 
(see the appendix).

\subsection{Subsonic MHD Flows}

Subsonic MHD flows are represented by Models~B1 and B2 (see Table~1)
and a have dynamically important magnetic field $\beta\sim 1-10$.
This field is injected into the computational domain in the form
of magnetic loops stretched in the radial direction (see Fig.~2).
The strength of the field is determined by the parameter $\beta_0$
(see Table~1 and eq.~[11]).
The total energy is conserved in Model~B1, whereas
the energy released in magnetic reconnections is completely lost
in Model~B2. 

Simulations of Models~B1 and B2 have been initiated from the 
steady hydrodynamic Model~A1 by gradually increasing the
strength of the injected magnetic field
to provide a smooth transition from nonmagnetic to magnetic flows.
Even in this case, however, Models~B1 and B2 have been 
settled into their final quasi-steady states after passing through
the initial transient phases.
The final quasi-steady states are turbulent and
characterized by random variations of all quantities 
on all spatial scales 
from the grid size to the size of the computational domain.
The amplitude of these variations, however, is not large and 
the time-averaged properties of the accretion flows remain unchanged 
on the large time scale.
The turbulence in Models~B1 and B2 is clearly originated because of
the effects of the magnetic field.

Figures~5--9 illustrate the structure of conservative Model~B1, showing
the snapshots of two-dimensional distributions of selected quantities 
in the equatorial cross section of the three-dimensional computational domain.
Figure~5 represents the density distribution, which shows
clearly recognizable small-scale (with respect to the scale $R$) 
density variations of the relative amplitude $\simeq 2$.
The lower and higher density regions typically take the form of filaments,
which are predominantly extended in the radial direction.
These radially extended structures can also be seen in
the velocity streamlines in Figure~6.
The streamlines form a complicated
flow pattern consisting of the radially extended narrow inflowing/outflowing
streams and small-scale vortices.

The time-dependent flow pattern in Model~B1 
demonstrates the randomly repeating events of the interchange instability.
In these events, colder, denser matter is
accumulated above (i.e., far from the center)
the region with hotter, lower density matter.
With time, this denser matter begins to move down through the low-density
region, forming a characteristic radially inflowing dense stream.
Such a stream typically forms a ``mushroom" at its head and
propagates about half a radius inward from the radius of origin.
At the same time, the stream carries
a frozen-in magnetic loop whose field strength is amplified
because of a radial convergence of the stream.
The example snapshot of magnetic lines in Figure~7 shows
many such magnetic loops resulting from the interchange instability.
During the subsequent evolution, the inflowing radial streams 
are fragmented into small pieces. 
This fragmentation is typically triggered by
reconnections of the oppositely directed 
magnetic lines confined in the streams.
The reconnections locally release energy and produce 
a hot low-density matter. This matter has
a positive buoyancy and forms narrow low-density outflows.

Figure~8 shows the snapshot of the plasma $\beta$.
The spatial distribution of $\beta$ and, accordingly, magnetic field
is highly nonuniform. The small-scale, high-$\beta$ regions, $\beta\ga 100$,
cover the entire computational domain and correspond to 
the weak field regions, which are typically associated with
reconnection regions. The large number of these regions clearly indicates
the high efficiency of the reconnection dissipation.
The low-$\beta$ regions, $\beta\la 1$, or, accordingly, 
the strong field regions are typically elongated in the radial direction and
associated with the inflowing streams. The field in such regions
becomes especially strong,
$\beta\ll 1$, in the innermost part of the flow, $R\la 2 R_{\rm in}$,
where the high-velocity inflowing streams dominate in the flow pattern.

The chaotic inward and outward turbulent motions 
in Model~B1 are only partially supported by magnetic interactions,
whereas the more important support is provided by thermal convection. 
As discussed earlier, magnetic reconnections 
are very efficient in the considered model and 
produce hot, low-density, narrow outflows.
Sometimes several such outflows can coalesce and form
a large convective bubble, which moves outward easier and faster.
The representative example of such a bubble can be seen
in Figure~5 as the low-density region near the outer boundary 
and, in Figure~9, as the corresponding increase of the specific entropy.
Later in time, this particular bubble has escaped through $R_{\rm out}$.
Figure~9 also shows many other local regions of high specific entropy. 
These regions typically coincide with the lower density
regions in Figure~5, have positive velocities, 
and can therefore be identified as convective bubbles. 
However, not all such bubbles will be able to escape through $R_{\rm out}$.
A significant part of these bubbles
will be pulled inward, mixed with the inflowing cold matter, and 
absorbed at $R_{\rm in}$.

To make the argument concerning the development of convection 
more quantitative,
we have calculated a one-dimensional time- and
angle-averaged distribution of the specific entropy in Model~B1
(see the description of the averaged procedure below).
This distribution has the negative slope and
therefore satisfies the Schwarzschild criterion for convection. 

The nonconservative Model~B2 is similar in many respects to 
the conservative Model~B1.
Model~B2 also forms a turbulent quasi-steady flow pattern.
Like in Model~B1, the developed nonuniformities in the density
frequently take the
form of narrow filaments, which are extended in the radial direction.
However, the amplitude and intensity of turbulent fluctuations in Model~B2
are significantly reduced in comparison with Model~B1.
As the result, the rate of reconnection dissipation is smaller in Model~B2 and
the average magnetic field is stronger.
The turbulent motions in Model~B2 involve different MHD effects and
are mainly supported by magnetic reconnections, magnetic buoyancy, and
interchange instability.
Convection motions are not developed in this model because of 
the absence of reconnection heat. The latter explains
the less efficient turbulence observed in this model.

Our MHD models are time variable
and, therefore, it is more practical
to study the structure of these models by
performing time and space averaging.
In this way, we have constructed one-dimensional distributions of different 
quantities, averaging them over the $\theta$ and $\phi$ directions. 
We have also averaged
the obtained spatially averaged distributions
over time, assuming the time-averaged interval $\tau\simeq 3 t_{\rm ff}$.
Figure~3 shows the averaged radial profiles of selected quantities
in Models~B1 and B2 (solid and dotted lines, respectively).
A comparison of these profiles shows certain differences 
in all quantities except in the accretion rates 
(see lower right panel in Fig.~3).
These rates are close to each other and almost independent of the radius. 
The latter confirms that the considered flows are in quasi-steady states.

Model~B2 shows a little steeper average
profiles for $\rho$ and $P_{\rm g}$ and a flatter profile of $v$
in the inner and middle radial ranges, $R \la 5 R_{\rm in}$, 
in comparison with Model~B1. 
The profiles of $P_{\rm m}$ show almost the same slopes in both models, but 
the magnitude of $P_{\rm m}$ is by the factor of $\simeq 2$ larger
in Model~B2 than in Model~B1.
As discussed earlier, this difference in $P_{\rm m}$ 
can be explained by the less
efficient dissipation of magnetic field in Model~B2.

The presence and absence of the reconnection heat resulted in different
temperatures in Models~B1 and B2 (see left lower panel in Fig.~3). 
In agreement with a naive expectation,
the conservative Model~B1 has the larger temperature.
At the same time, this model has a larger magnetic Mach number, 
${\cal M}_{\rm m}= v/\sqrt{c_s^2+c_{\rm A}^2}$, where
$c_{\rm A}=\sqrt{B^2/4\pi\rho}$ is the Alfvenic speed, and larger 
$v$ in the inner region
(see right top and right middle panels in Fig.~3).
The latter two facts can not be simply understood using the 
analogy with Bondi-type flows in which lower ${\cal M}$ and $v$
correspond to hotter flows.  It seems that the stronger magnetic field
is responsible for the relative reduction of these quantities in Model~B2.

The averaged radial structures of MHD Models~B1 and B2 show noticeable 
differences from the radial structures of hydrodynamic Model~A1 
and self-similar solution (A9) (see Fig.~3).
The MHD flows have the flattened density profiles.
In terms of the power-law distribution $\rho\propto R^{-\sigma}$,
Models~B1 and B2 have $\sigma\approx 0.7$, whereas
Model~A1 and solution (A9) have respectively
$\sigma\approx 1.3$ and $\sigma=1.5$ (see left upper panel in Fig.~3).
Similarly, Models~B1 and B2 have the flattened profiles of
$P_{\rm g}\propto R^{-\eta}$ in which $\eta\approx 1.5$, whereas
Model~A1 and solution (A9)  have $\eta\approx 2.2$ 
and $\eta=2.5$, respectively (see left middle panel in Fig.~3).
Other differences between hydrodynamic and MHD flows
include the larger temperatures and reduced
accretion velocities in the latter flows
(see left lower and right middle panels in Fig.~3).
The accretion rates in Models~B1 and B2 show an $\simeq 10$\% reduction
with respect to Model~A1 (see right lower panel in Fig.~3).
This is a clear indication that turbulent MHD accretion flows
have reduced accretion rates.

\subsection{Supersonic MHD Flows}

Supersonic MHD flows are represented by Models~C1 and C2 (see Table~1). 
These models have been initiated from the supersonic hydrodynamic Model~A4 
by injecting a loop magnetic field at $R_{\rm inj}$.
Similar to the subsonic MHD models (see \S 3.2), 
Models~C1 and C2 have approached
quasi-steady states after passing through the transient phases.
However, the transient phases in these models show some differences:
the developed temporal shocks and waves are downshifted
by the supersonic inflow and never reached the outer boundary $R_{\rm out}$.
In quasi-steady states, each of these models obtains a new 
feature---a quasi-steady shock.
This shock divides the flows into two regions:
the outer and inner, which have super-fast and
sub-fast-magnetosonic accretion velocities, respectively.
In the outer region, the flows are radial and laminar.
The magnetic field is dominated by the radial component; this
is quickly increased inward as $B\propto R^{-2}$. No significant
field dissipation and reconnections have been observed in this region.
In the inner region, on the other hand,
reconnections become important and the flow is turbulent. 
The quasi-steady shocks in both Models~C1 and C2 can be classified 
as Alfvenic shocks; they change the orientation of magnetic lines, 
but leave the field strength almost unchanged. 
Right after the completion of the transient phases, 
these shocks have been located 
at $R\simeq 2 R_{\rm in}$ in both Models~C1 and C2.

During the subsequent evolution, Models~C1 and C2
have been slowly evolved on the time scales $\gg t_{\rm ff}$. 
As a result of this evolution,
the inner regions in these models have been expanded, and, consequently,
the Alfvenic shocks have slowly moved outward.
The shock moves relatively faster in the case of the conservative Model~C1
and slower in the case of the nonconservative Model~C2.
Specifically, during the evolution time $t\simeq 40 t_{\rm ff}$ 
counted from the end of the transient phases, 
the shock has propagated a distance of $\simeq 6\, R_{\rm in}$
in Model~C1 and the distance $\simeq R_{\rm in}$ in Model~C2. 
It is worth noting that these slowly moving Alfvenic shocks
do not have analogy in hydrodynamic accretion flows.
In the latter flows, any developed shocks are nonstationary and
move either inward or outward, depending on the assumed conditions, 
on the timescale of $\sim t_{\rm ff}$.\footnote{ 
We do not consider the radiative shocks here, which can be (quasi) steady.}

Figure~10 illustrates the flow pattern in the conservative Model~C1, showing
the velocity streamlines projected onto the equatorial plane.
The Alfvenic shock is located at $R\approx 7\, R_{\rm in}$ in the shown moment.
Two flow regions can be clearly distinguished: the pre-shock
laminar outer and post-shock turbulent inner regions. Figure~11 shows 
magnetic lines that correspond to the flow pattern shown in Figure~10.
In the pre-shock region, magnetic lines are purely radial, but
oppositely directed in different sectors (see Fig.~2). 
These lines do not experience reconnections 
because of the purely radial flow pattern.
In the post-shock region, magnetic lines are tangled because of the turbulent
flow pattern and frequently reconnected.

The difference in Alfvenic shock motions in Models~C1 and C2 
suggests that there are two different mechanisms that cause these motions.
Model~C1, which includes the reconnection heat,
develops a convection in the post-shock region, 
similar to the convection in Model~B1.
This convection transports the heat outward, causing
the relatively fast expansion of the inner region and
faster motion of the shock.
This mechanism, however, is not suitable for Model~C2, which does not
include the reconnection heat and does not develop convection.
Instead, the shock motion in Model~C2 can be explained by
magnetic buoyancy, which acts in the post-shock region
in a manner similar to the action of magnetic buoyancy in Model~B2.
As a result, the magnetic field and energy
are transported outward, forcing the inner region to expand 
and move the shock outward.
The slower expansion of the post-shock in Model~B2 is explained by
the less-efficient energy transport provided by magnetic buoyancy.

The radial distributions of selected quantities in Models~C1 and C2 are 
shown in Figure~4 by solid and dotted lines, respectively. 
These distributions have been
obtained employing the averaging techniques described in \S 3.3
in which the time-averaged interval is chosen to be smaller,
$\tau\simeq 0.1 t_{\rm ff}$,
to avoid ``washing out" the moving shocks.
One can clearly see that
the structure of Models~C1 and C2 changes quite sharply
at the Alfvenic shocks, whose locations at the shown moments are
$R \approx 7.5\, R_{\rm in}$ and
$\approx 3\, R_{\rm in}$, respectively.

The outer regions in Models~C1 and C2 are
almost identical to each other and similar to the outer part of
Model~A4 (the dashed lines in Fig.~4). 
This can be seen in the distributions of all quantities except for the 
magnetic Mach number ${\cal M}_{\rm m}$ (see right upper panel in Fig.~4). 
Models~C1 and C2 include strong magnetic fields that significantly reduce
${\cal M}_{\rm m}$
(but still ${\cal M}_{\rm m} > 1$ before the shocks), whereas the accretion
velocities remain almost the same as in hydrodynamic Model~A4  
(see right middle panel in Fig.~4).
Note the interesting behavior of $P_{\rm g}$ and $P_{\rm m}$
in these models (see left middle panel in Fig.~4). 
The values of $P_{\rm g}$ before the shocks in the MHD models closely follow 
the corresponding value in Model~A4.
The values of $P_{\rm m}$, which are started from the subequipartition 
level at $R_{\rm inj}$, $\beta=10$,
quickly exceed the equipartition level, 
increasing inward as $P_{\rm m}\propto R^{-4}$.
In the case of Model~C2, $\beta\approx 0.1$
just before the shock and the energy of the flow is dominated by 
the kinetic and magnetic energies, which are of approximately equal magnitudes.

The inner turbulent regions in Models~C1 and C2 are 
sub-fast-magnetosonic, ${\cal M}_{\rm m} < 1$, and have Alfvenic or moderately
super-Alfvenic accretion velocities, $v \ga c_a$.
These regions are more dense,
more hot, and have lower accretion velocities than
the corresponding part of Model~A4 (see Fig.~4).
Magnetic reconnections are important here; they
support turbulence and provide dissipation of the magnetic fields.
This dissipation noticeably reduces the slope of $P_{\rm m}$ 
with respect to the corresponding slope
in the outer region of the flow where the dissipation is negligibly small.
In the case of Model~C2,
the power-law index $\eta$ for the magnetic pressure distribution 
$P_{\rm m}\propto R^{-\eta}$ 
is changed from $\eta=4$ in the outer part
to $\eta\approx 3$ in the inner part of the flow
(see left middle panel in Fig.~4).

\section{Discussion and Conclusions}

We have numerically investigated quasi-steady MHD spherical accretion flows
with imposed small-scale magnetic fields. We have confirmed the previous
theoretical expectation and numerical results such that
the flows are turbulent and have radial structures different
from the Bondi-type accretion flows (e.g., Shvartsman 1971; IN).
Turbulence in the MHD flows is developed and supported by
the interchange instability, thermal convection,
and various magnetic interactions, 
including magnetic reconnections and buoyancy. 
The highly nonuniform release
of energy in reconnections, which is, in fact, the release of 
the gravitational energy converted and stored in the form of magnetic field, 
makes these flows so special and different
from the laminar and stable Bondi-type flows (however, 
see Kovalenko \& Eremin 1998).
We have found that magnetic buoyancy, in addition to thermal convection,
can play an important role
in the modification of the flow structure, especially in the case
of nonconservative (or radiatively efficient) flows 
in which the convection could not be developed.

The most important consequence of turbulence in our MHD
models, both conservative and nonconservative, 
is the modification of the radial flow structure. 
Figure~3 demonstrates this modification.
Turbulent subsonic MHD Models~B1 and B2 have the flattened time-averaged
density profiles, higher temperatures, and lower accretion velocities 
in comparison with their hydrodynamic counterpart Model~A1. 
These properties make Models~B1 and B2 more like spherical CDAFs
(see IN) than Bondi accretion flows.
The theory of spherical CDAFs predicts the flattened power-law density profile 
$\rho\propto R^{-\sigma}$ in which $\sigma=0.5$. Models~B1 and B2 have
profiles close to this, but more steep, $\sigma\approx 0.7$.
These steeper profiles can be explained by the
influence of the inner boundary condition
in our numerical models, whereas the analytic CDAF solution 
is ``boundary free."
Abramowicz {\it et al.} (2002) argued
that the proximity of the black hole absorbing boundary makes the
inner regions of CDAFs to be advection dominated.
In particular, they demonstrated that viscous rotating CDAFs
become advection dominated inside
$\sim 50$ to 100 gravitational radii.
Our numerical models have a rather small radial range,
$R_{\rm out}/R_{\rm in}=10$, and, therefore,
the effects of the inner boundary can be significant.

Turbulent Models~B1 and B2 have shown the
reduction of the accretion rates in comparison with laminar 
Model~A1 (see right lower panel in Fig.~3). 
This result qualitatively confirms
the predicted reduction of the accretion rate in
spherical CDAFs (see eq.~[1]). 
However, the actual reduction of the accretion rates, 
which is about 10\%, 
shows a poor quantitative agreement with estimate (1). 
We explain this poor agreement by
the limited radial range in our models, whereas
estimate (1) was obtained in the limit $R_{\rm out}\gg R_{\rm in}$.

The conservative and nonconservative subsonic MHD models 
(Models~B1 and B2, respectively) are turbulent, 
however, the turbulence properties in these models are different. 
In the conservative model, the turbulent motions are more intensive and
mainly supported by thermal convection, which makes this model
similar to spherical CDAFs. The nonconservative model
has less intensive turbulence, which is supported through
various magnetic interactions in which magnetic buoyancy seems
to be dominated.
In some respect, the magnetic buoyancy acts similar to the thermal convection;
it also transports (magnetic) energy outward. This transport 
can explain the flattened density profile in the nonconservative model,
like the convection in spherical CDAFs (see IN). 
However, the energy transport provided by magnetic buoyancy is less
efficient and much weaker than the convection transport. 

Our supersonic MHD accretion flows (Models~C1 and C2) 
have been initiated by injecting a low
entropy matter into the computational domain. 
These flows form quasi-steady Alfvenic shocks that separate the laminar 
super-fast-magnetosonic outer inflows and the post-shock turbulent 
nearly Alfvenic inner inflows. 
We have found that the averaged flow pattern in these models is not
steady on the large time scale $\gg t_{\rm ff}$. 
The post-shock regions gradually
expand, forcing the Alfvenic shocks to move outward. These
regions are quite similar to
the subsonic MHD flows in Models~B1 and B2, and, therefore, the gradual
expansion of these regions can be explained by the
outward energy flux provided by convection and/or magnetic buoyancy.
We have found that the Alfvenic shock moves significantly faster
in the case of Model~C1, which develops a convection in 
the post-shock region.
The latter is consistent with our observation that the 
more intensive turbulence and, respectively, the larger outward energy flux
happens in convective Model~B1, rather than in nonconvective Model~B2.

Based on our study of the supersonic MHD models,
we conclude that {\it stationary} supersonic (or super-fast-magnetosonic)
accretion flows cannot be realized
in the presence of small-scale magnetic fields.
This conclusion is equally applied to radiatively efficient and
inefficient flows.
Such supersonic flows will unavoidably create shocks at the
equipartition radius and these shocks will be moved outward 
because of the action of
convection and magnetic buoyancy, which are developed in turbulent
post-shock regions of these flows.
These post-shock regions will continue to expand and,
on the large time scale, fill the entire accretion domain,
causing the flows to be subsonic everywhere.

The wedge computational domain of our models has the limited
opening angle (see Fig.~1), 
and we use the specific boundary conditions in the angular directions
(see Section~2.1) to minimize the influence of this limited-size domain.
Definitely, the used boundary conditions could have some effects
on the properties of the simulated MHD turbulence, for example, 
limiting the spatial scales and effecting 
the motions in the vicinity of the polar sliding boundaries.
Other consequence of the employing this domain is the inability
to simulate large scale magnetic structures, which can be developed
during the reverse cascade of energy from small to large spatial scales 
in a MHD turbulence. This issue of the limited computational domain
should be addressed in future works.

The accretion of mass in our MHD models is accompanied by
the reconnection dissipation of the magnetic field in turbulent flows. 
The rate of dissipation is consistently regulated through feedback 
mechanisms that also regulate the intensity of the turbulence. 
To illustrate the dissipation process quantitatively,
we consider the nonconservative MHD models in which the change
of the time-averaged radial flux of the total energy
\begin{equation}
F_R=\int_{\Omega_0} R^2 q_R d\Omega
\end{equation}
directly corresponds to the amount of dissipated magnetic energy.
The integral in equation (13) is taken over the solid angle 
of the computational domain $\Omega_0$ and 
the radial flux per unit square (see eq.~[3]) is
\begin{equation}
q_R=\rho v_R\left({v^2\over 2}+\epsilon +{P\over\rho}
+{B^2\over 4\pi\rho}-{GM\over R}\right)-
{B_R\over 4\pi}({\bf v}\cdot {\bf B}).
\end{equation}
The solid lines in Figure~12 represent the radial dependence
of the total energy fluxes obtained in Models~B2 and C2. 
These fluxes have been normalized to the flux
$GM\dot{M}_{\rm in}/R_{\rm in}$. The dashed lines in Figure~10
correspond to the flux conservation $F_R={\rm const}$.
The difference between the dashed and solid lines 
at a given $R$ represents the amount of the energy dissipated
in the radial range from $R_{\rm out}$ to $R$. Note that
the energy dissipates in the whole volume in Model~B2,
whereas it dissipates only in the post-shock
region and no energy dissipates in the pre-shock region in Model~C2.
The total amount of reconnection losses given
in units of gravitational energy at $R_{\rm in}$
is about 6.5\% in Model~B2 and about 5\% in Model~C2.

\acknowledgments

The author gratefully thanks Ramesh Narayan and Igor Novikov for 
discussions, and Vasilij Beskin for usefull comments.
This work was supported by
the U.S. Department of Energy (DOE) Office of Inertial Confinement
Fusion under Cooperative Agreement No. DE-FC52-92SF19460, the
University of Rochester, the New York State Energy Research and
Development Authority.

\appendix

\section{Some Analytic Solutions of Spherical Hydrodynamic Accretion Flows}

In this paper, we have compared our numerical models and
analytic solutions of spherical hydrodynamic accretion flows.
We reproduce some relevant formulas for these solutions below.

Spherical, stationary, and isentropic hydrodynamic accretion flows
are described by the continuity equation
\begin{equation}
\dot{M}=4\pi R^2 \rho v,
\end{equation}
Bernoulli's equation
\begin{equation}
{v^2 \over 2}+{\gamma\over\gamma-1}{P\over\rho}-{GM\over R}={\rm constant},
\end{equation}
and the politropic equation of state
\begin{equation}
P=K\rho^\gamma,
\end{equation}
where $P$ is the gas pressure,
$K$ is the politropic constant, $\gamma$ is the politropic index,
and $\dot{M}$ is the accretion rate.
Bondi (1952) solved these equations assuming that the flow
is at rest and of the uniform density $\rho_\infty$,
pressure $P_\infty$, and sound speed
$c_{\infty}=\sqrt{\gamma P_\infty/\rho_\infty}$ at infinity.
The Bondi solution depends on the spatial scale
$R_B=2GM/c_{\infty}^2$, called the Bondi or accretion radius, and
can be expressed in the following implicit form that
gives the radial dependence of $\rho$:
\begin{equation}
\lambda^2\left({R_B\over R}\right)^4
\left({\rho_\infty\over \rho}\right)^2+
{2\over\gamma-1}\left[\left({\rho\over \rho_\infty}\right)^{\gamma-1}
-1\right]-{R_B\over R}=0,
\end{equation}
where $\lambda=\dot{M}/(4\pi R_B^2\rho_\infty c_{\infty})$ is the
dimensionless accretion rate defined by the expression
\begin{equation}
\lambda=\left({1\over 2}\right)^{5\gamma-3\over 2(\gamma-1)}
\left({5-3\gamma\over 4}\right)^{3\gamma-5\over 2(\gamma-1)}.
\end{equation}
In the small-radii limit, $R \ll R_B$, the Bondi solution (A4)
is represented by the asymptotic power-law distributions
\[
v\propto R^{-1/2},
\]
\begin{equation}
\rho\propto R^{-3/2},
\end{equation}
\[
P\propto R^{-5/2}.
\]
In the case of $1\le\gamma < 5/3$, the
Bondi solution (A4) describes transonic flows in which the Mach number
\begin{equation}
{\cal M}=\lambda\left({R_B\over R}\right)^2
\left({\rho_\infty\over \rho}\right)^{(\gamma+1)/2}
\end{equation}
is zero at infinity and 
monotonically increases inward, approaching the infinite value at $R=0$.

In the subsequent discussion, we shall consider only the case $\gamma=5/3$, 
which is known to be the special case of the Bondi solution (A4).
In this case, the flow is subsonic everywhere and ${\cal M}=1$ only
in the limit of $R\rightarrow 0$. However, such a ``boundary free"
solution is not practical for numerical applications and
we have modified the Bondi solution to include the finite inner boundary
radius $R_{\rm in}>0$ in which we require ${\cal M}=1$.
The modified Bondi solution is described by same equation (A4)
and the same equation (A7) represents the Mach number, but a new
$\lambda$ is defined as follows:
\begin{equation}
\lambda=\left({1+3\delta\over 4}\right)^2,
\end{equation}
where $\delta=R_{\rm in}/R_B$ is a free parameter; $0 < \delta < 1$. 
This modified solution approximates well our subsonic Model~A1,
assuming $\delta=0.00371$ (short-dashed lines in Fig.~3).

In the case of $\gamma=5/3$, equations (A1)-(A3) allow
a self-similar solution, which requires the special outer
boundary condition $c_{\infty}=0$ or, equivalently, to set
the constant on the right hand side of equation (A2) to zero.
This solution reads
\[
v=\alpha\sqrt{GM/R},
\]
\begin{equation}
\rho=\dot{M}/( 4\pi\alpha\sqrt{GM}R^{3/2}),
\end{equation}
\[
P={2\over 5}\left(1-{\alpha^2\over 2}\right){GM\rho\over R},
\]
where $\alpha$ is a free parameter; $0 < \alpha < \sqrt{2}$.
Solution (A9) can be subsonic or supersonic depending on $\alpha$. 
This solution is characterized by the constant Mach number
\begin{equation}
{\cal M}=\alpha\sqrt{1.5/(1-\alpha^2/2)},
\end{equation}
and it is supersonic at $1/\sqrt{2} < \alpha < \sqrt{2}$.

Another kind of supersonic solution in the special case of $\gamma=5/3$
can be constructed assuming the outer boundary 
at some finite radius $R_{\rm out}$ in which we require ${\cal M}=1$.
This solution takes the form
\begin{equation}
\tilde{\lambda}^2\left({\tilde{R}_B\over R}\right)^4
\left({\rho_{\rm out}\over \rho}\right)^2+
3\left[\left({\rho\over \rho_{\rm out}}\right)^{2/3}
-1\right]-{\tilde{R}_B\over R}+{1\over\tilde{\delta}}-1=0,
\end{equation}
where $\tilde{R}_B=2GM/c_{\rm out}^2$ and 
$\tilde{\delta}=R_{\rm out}/\tilde{R}_B$, and we denote $\rho_{\rm out}$ 
and $c_{\rm out}$ to be the density
and sound speed at $R_{\rm out}$, respectively.
Solution (A11) exists for $0 < \tilde{\delta} \le 1/4$.
The dimensionless accretion rate
$\tilde{\lambda}=\dot{M}/(4\pi \tilde{R}_B^2 \rho_{\rm out} c_{\rm out})$
can be expressed in the form
\begin{equation}
\tilde{\lambda}=\tilde{\delta}^2.
\end{equation}
The Mach number in solution (A11)
\begin{equation}
{\cal M}=\tilde{\lambda}\left({\tilde{R}_B\over R}\right)^2
\left({\rho_{\rm out}\over \rho}\right)^{4/3}
\end{equation}
monotonically increases inward and
approaches the asymptotic value ${\cal M}_0$ at $R\ll R_{\rm out}$.
In the limit $\tilde{\delta} \ll 1$, one gets
${\cal M}_0=\tilde{\delta}^{-2/3}$.
In the case of marginal $\tilde{\delta}=1/4$, 
one gets ${\cal M}={\cal M}_0=1$ and solution (A11) becomes equivalent 
to the self-similar solution (A9) in which $\alpha=1/\sqrt{2}$.
Solution (A11) approximates well our supersonic Model~A4,
assuming $\tilde{\delta}=0.071$ (short-dashed lines in Fig.~4).

\clearpage

% ---- TABLE 1 -----
\begin{deluxetable}{cccccc}
\footnotesize \tablecaption{Parameters of the Models \label{tbl-1}}
\tablewidth{0pt} \tablehead{ \colhead{Model} &
\colhead{~~~~$\epsilon_0$\tablenotemark{a}~~~~} &
\colhead{~~~~$\beta_0$\tablenotemark{b}~~~~} &
\colhead{~Reconnection heat~~} &
\colhead{Field topology}}
\startdata A1 & 1.2 & 0      & ---  & ---   \nl
           A2 & 1.2 & $10^2$ & ---  & Toroidal \nl
           A3 & 1.2 & 30     & ---  & Radial \nl
           A4 & 0.3 & 0      & ---  & ---  \nl
%          subsonic models
           B1 & 1.2 & 10     & Included & Loop\nl
           B2 & 1.2 & 10     & Not included  & Loop\nl
%          supersonic models
           C1 & 0.3 & 10     & Included & Loop\nl
           C2 & 0.3 & 10     & Not included  & Loop\nl
\enddata
\tablenotetext{a}{Characterizes the specific internal energy of
the injected mass given in units of $\epsilon_{\rm vir}$ (see eq.~[5]).
The flows with $\epsilon_0=1.2$ are subsonic and $\epsilon_0=0.3$ 
are supersonic.}
\tablenotetext{b}{Characterizes the initial or injected
magnetic field strength and is defined in eq.~(11).}
\end{deluxetable}

\clearpage

\begin{figure}
\epsscale{.70}
\plotone{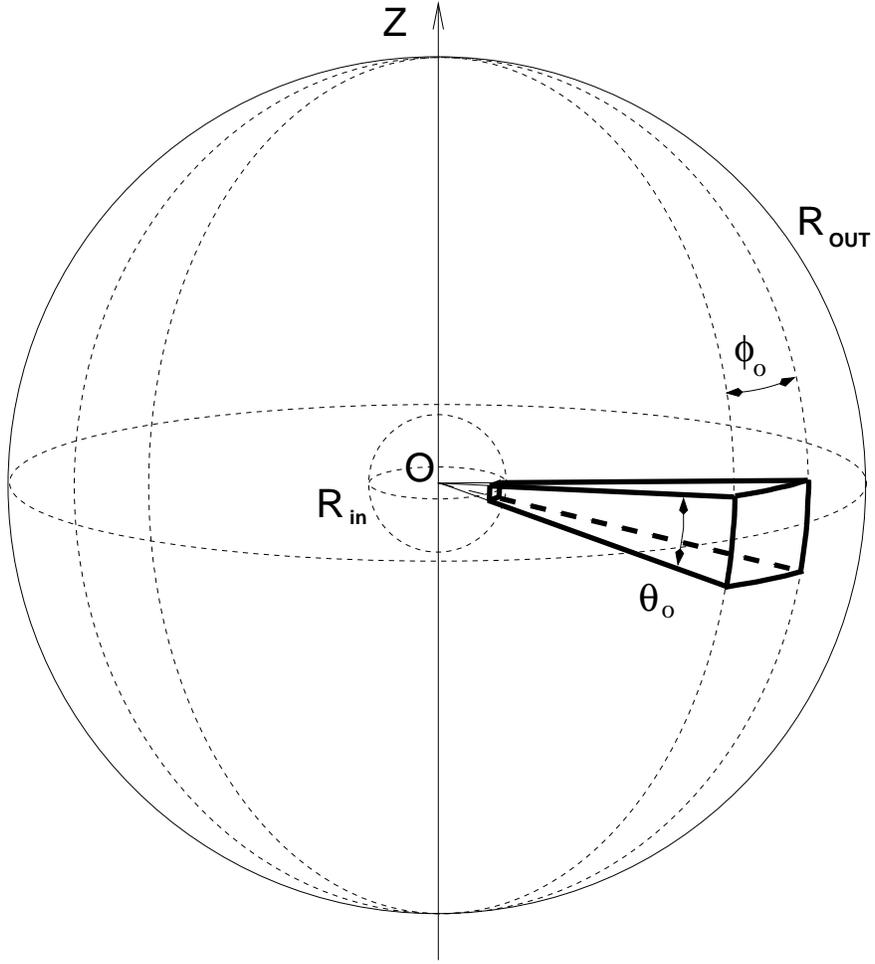}
\caption{Geometry of the computational domain used in the simulations.
The domain is extended from $R_{\rm in}$ to $R_{\rm out}$ 
in the radial direction ($R_{\rm out}/R_{\rm in}=10$) 
and spans the angles $\theta_0$ and $\phi_0$
in the polar and azimuthal directions, respectively
($\theta_0=\phi_0=\pi/16$).
The source of gravity is located in the origin.
\label{fig1}}
\end{figure}

\clearpage

\begin{figure}
\epsscale{.60}
\plotone{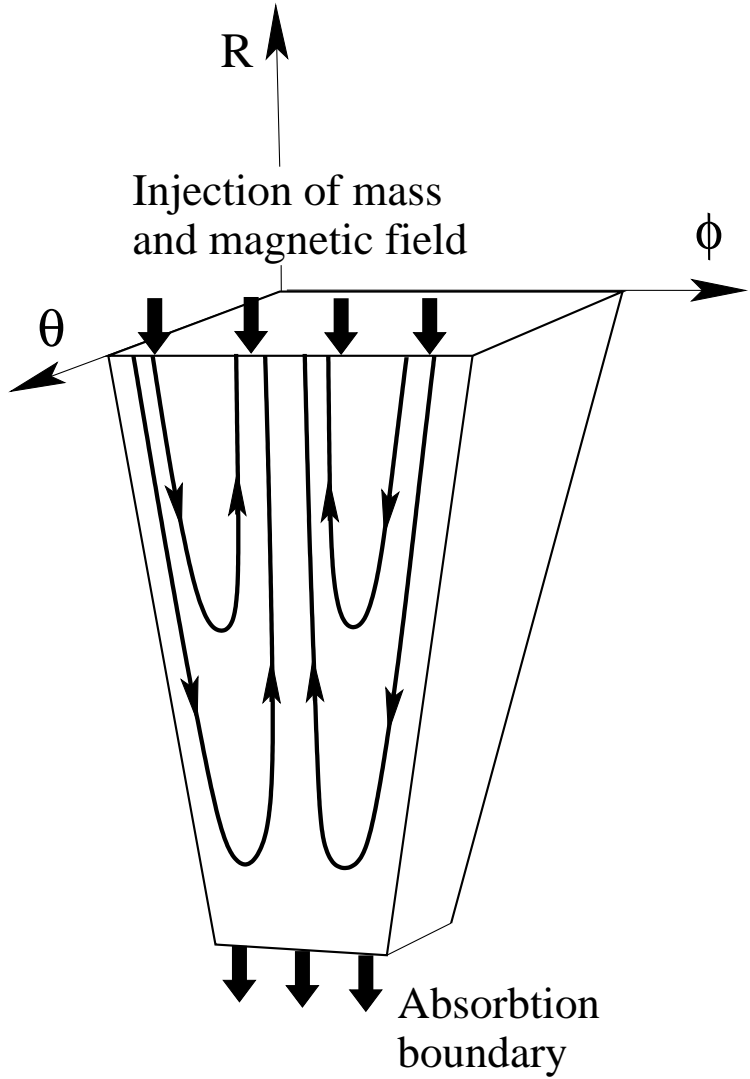}
\caption{Schematic illustration of the simulation design. 
A three-dimensional MHD accretion flow 
is formed as the result of a steady injection of
mass and magnetic field in the vicinity of the outer radial boundary
in the wedge computational domain. The geometry of the injected field
is represented by magnetic lines (thick solid lines with arrows).
The accretion flow with a frozen-in magnetic field is absorbed
at the inner radial boundary.
\label{fig2}}
\end{figure}

\clearpage

\begin{figure}
\plotone{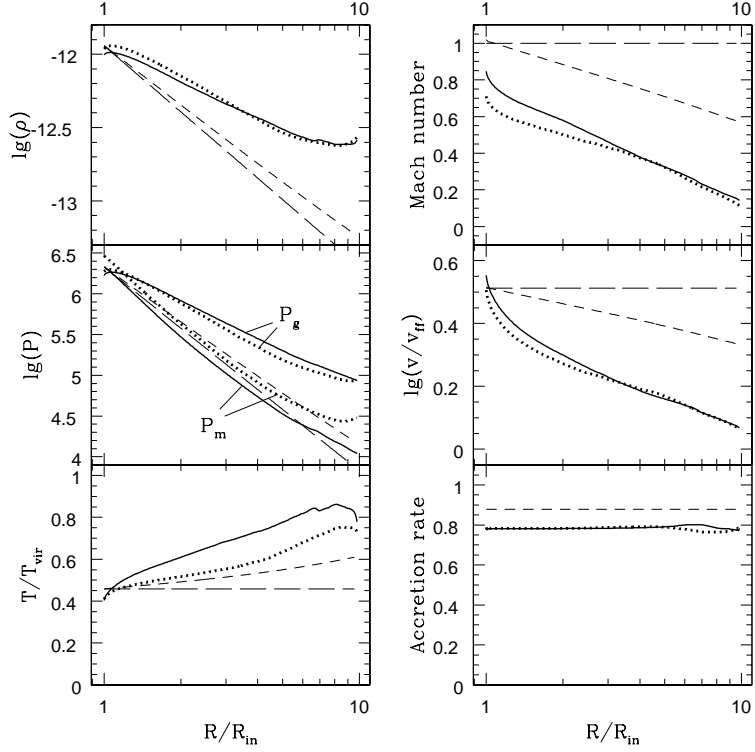}
\caption{Radial structure of the subsonic accretion flows in Models~A1 
(short-dashed lines),
B1 (solid lines), B2 (dotted lines), and self-similar solution
(long-dashed lines, see eq.~[A9]). 
Hydrodynamic Model~A1 is steady; turbulent MHD Models~B1 and B2 are
shown in quasi-steady states.
All plotted quantities in the turbulent models---the 
density $\rho$, gas and
magnetic pressures $P_{\rm g}$ and $P_{\rm m}$, temperature $T$,
magnetic Mach number ${\cal M}_{\rm m}$, accretion velocity $v$, 
and mass accretion
rate $\dot{M}$---have been averaged over the angles $\theta$ and $\phi$
and over the time interval $\tau\simeq 3 t_{\rm ff}$.
\label{fig3}}
\end{figure}

\clearpage
                                                                                
\begin{figure}
\plotone{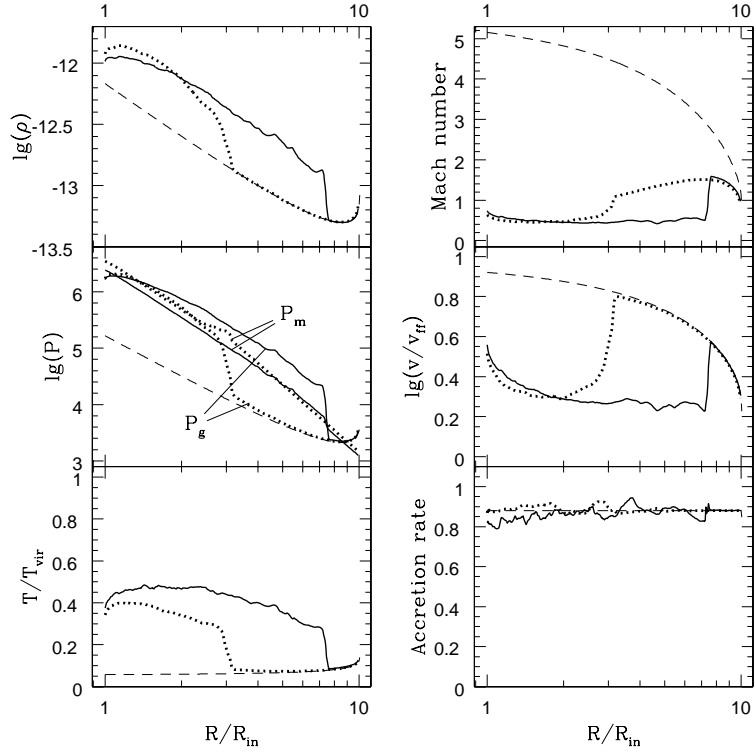}
\caption{Radial structure of the supersonic accretion flows in Models~A4 
(short-dashed lines), C1 (solid lines), and C2 (dotted lines).
Hydrodynamic Model~A4 is steady; turbulent MHD Models~C1 and C2 are
shown in quasi-steady states.
All plotted quantities in the turbulent models---the density $\rho$, gas and
magnetic pressures $P_{\rm g}$ and $P_{\rm m}$, temperature $T$,
magnetic Mach number ${\cal M}_{\rm m}$, accretion velocity $v$, 
and mass accretion
rate $\dot{M}$---have been averaged over the angles $\theta$ and $\phi$
and over the time interval $\tau\simeq 0.1 t_{\rm ff}$.
The MHD models have slowly moving outward Alfvenic shocks, which
are located at the shown moment at $R\approx 7.5 R_{\rm in}$
in Model~C1 and $R\approx 3 R_{\rm in}$ in Model~C2.
\label{fig4}}
\end{figure}

\clearpage

\begin{figure}
%\plotone{dens_dcc37.eps}
%\plotone{f5.eps}
\caption{Snapshot of the density distribution in Model~B1 
in the equatorial plane. The source of gravity is located
on the left, outside of the shown cross-section of the computational domain
(for a general view see Fig.~1).
The shown radial range $R_{\rm out}/R_{\rm in}=10$ represents the
fully simulated domain.
The mass and magnetic field are steadily injected near the outer
radial boundary on the right.
The color bar on the left indicates the scale
for $\log\rho$ (in arbitrary units).
The model is turbulent and shown in a quasi-steady state.
The fluctuations in the density are
caused mainly by nonuniform reconnection heat and 
supported by convection motions.
\label{fig5}}
\end{figure}

\begin{figure}
%\plotone{slines_bw_dcc37.eps}
%\plotone{f6.eps}
\caption{Snapshot of velocity streamlines in Model~B1 in the equatorial plane
in the same moment as in Fig.~5.
The component of the streamlines parallel to the plane is shown.
The flow pattern consists of the radially (horizontal in this view)
extended narrow inflowing/outflowing streams and small-scale vortices.
The streamlines have been plotted using the line integral
convolution method of Cabral \& Leedom (1993).
\label{fig6}}
\end{figure}

\begin{figure}
%\plotone{mlines_bw_dcc37.eps}
%\plotone{f7.eps}
\caption{Snapshot of magnetic lines in Model~B1 in the equatorial plane
in the same moment as in Fig.~5.
The component of the field lines parallel to the plane is shown.
One can see many magnetic loops stretched in the radial (horizontal
in this view) direction. These loops are the result of convection and
interchange instability.
\label{fig7}}
\end{figure}

\begin{figure}
%\plotone{beta_dcc37.eps}
%\plotone{f8.eps}
\caption{Snapshot of the distribution of the plasma 
$\beta\equiv P_{\rm g}/P_{\rm m}$ in Model~B1 in the equatorial plane
in the same moment as in Fig.~5.
The color bar on the left indicates the scale for $\log\beta$.
Small-scale regions of large $\beta$, or weak magnetic fields, correspond to
regions of reconnection and dissipation of magnetic energy.
The large number of these regions indicates the high efficiency
of the reconnection dissipation. 
Regions of low $\beta$, or strong magnetic fields, are typically
elongated in the radial (horizontal in this view) direction and
associated with the inflowing streams.
\label{fig8}}
\end{figure}

\begin{figure}
%\plotone{entr_dcc37.eps}
%\plotone{f9.eps}
\caption{Snapshot of the distribution of specific entropy $s$
in Model~B1 in the equatorial plane in the same moment as in Fig.~5.
The color bar on the left indicates the scale for $\log s$
(in arbitrary units).
Regions of large $s$ represent hot convective bubbles and streams,
which have positive buoyancy and typically move outward.
Regions of low $s$ correspond to relatively cold inflowing matter.
On average, the specific entropy is increased inward.
\label{fig9}}
\end{figure}

\begin{figure}
%\plotone{slines_bw_dbd05.eps}
%\plotone{f10.eps}
\caption{Snapshot of velocity streamlines in supersonic
Model~C1 in the equatorial plane.
The component of the streamlines parallel to the plane is shown.
The flow pattern consists of two regions: the laminar
super-fast-magnetosonic pre-shock (outer, on the right in this view)
region and turbulent/convective post-shock 
(inner, on the left in this view) region.
A quasi-steady Alfvenic shock separates these two regions and is located
at $R\approx 7\, R_{\rm in}$ in the shown moment.
This shock slowly moves outward.
\label{fig10}}
\end{figure}
                                                                                
\begin{figure}
%\plotone{mlines_bw_dbd05.eps}
%\plotone{f11.eps}
\caption{Snapshot of magnetic lines in supersonic 
Model~C1 in the equatorial plane
in the same moment as in Fig.~10.
The component of the field lines parallel to the plane is shown.
The field lines are purely radial 
(but oppositely directed in different sectors, see Fig.~2) 
and do not experience reconnections
in the laminar pre-shock region (see capture to Fig.~10).
The tangled field lines in the post-shock region are the result 
of convection and interchange instability.
\label{fig11}}
\end{figure}

%\clearpage

\begin{figure}
\epsscale{.70}
%\plotone{bernoulli.ps}
%\plotone{f12.eps}
\caption{Radial distribution of the time-averaged total energy flux $F_R$
normalized to the flux $GM\dot{M}_{\rm in}/R_{\rm in}$
in Models~B2 and C2 (solid lines). 
Reduction of $F_R$ with respect to its value at the outer boundary
(shown by dashed lines) represent losses of magnetic energy 
in reconnections. Supersonic Model~C2 is shown at the moment, which corresponds
to the Alfvenic shock position $R\approx 4.7\, R_{\rm in}$.
Note the absence of energy losses in the laminar
pre-shock region in Model~C2.
\label{fig12}}
\end{figure}

\end{document}